\documentclass[epj]{svjour}
\usepackage{epsfig}
\def\lsim{\raise0.3ex\hbox{$<$\kern-0.75em\raise-1.1ex\hbox{$\sim$}}}
\def\gsim{\raise0.3ex\hbox{$>$\kern-0.75em\raise-1.1ex\hbox{$\sim$}}}
\begin{document}
\title{Deconfinement and Quarkonium Suppression}
\author{Frithjof Karsch\inst{1,2}
\thanks{\emph{Present address:} Physics Department, Brookhaven National
Laboratory, Upton, NY 11973, USA}%
}                     
\institute{
Fakult\"at f\"ur Physik, Universit\"at Bielefeld,
D-33615 Bielefeld, Germany \and
Physics Department, Brookhaven National
Laboratory, Upton, NY 11973, USA}
\date{Received: date / Revised version: date}
%
\abstract{
Modifications in the production pattern
of heavy quark bound states have long been considered to provide sensitive
signatures for the thermal properties of dense matter created in heavy ion
collisions. The original concept of Matsui and Satz for quarkonium suppression 
as signature for deconfinement in heavy ion collisions has been challenged 
recently through lattice studies of spectral functions, which indicate the 
persistence of heavy quark bound states at temperatures well above the 
transition, as well as through the refined analysis of hadronization and 
recombination models, which take into account the thermal evolution of the 
medium generated in a heavy ion collision. We will review here recent 
developments on these topics. 
\PACS{{11.15.Ha},{ 11.10.Wx},{ 12.38.Mh},{ 25.75.Nq}}
} 
\maketitle
\section{Introduction}
\label{intro}
The concept of quarkonium suppression as a probe for the
thermal properties of hot and dense particle matter or, more specifically,
for the deconfining nature of the transition between hadronic matter at
low temperatures and a dense partonic medium at high temperatures
has originally been put forward by Matsui and Satz \cite{Matsui}. In their 
approach they utilized the long-distance infrared properties of QCD; in a 
deconfined medium the heavy quark potential gets screened and at sufficiently 
high temperature the screening radius will be smaller than the typical size
of a quarkonium state. As a consequence the screened potential no longer 
can support the formation of bound states.  

Screening strongly depends on a sufficiently high density of partons, 
{\it i.e.} color charges that can contribute to the screening. In a QED
plasma, for instance, the screening radius $r_D$ and density $n$ are related 
through $r_D \sim 1/g\sqrt{n/T}$. The rapid rise of the parton density at 
a well defined transition temperature, irrespective of this being a phase 
transition temperature or a temperature characterizing a rapid 
crossover, thus intimately connects the occurrence of a (phase) transition
in QCD with the rapid decrease of the screening radius and thus also with
the fate of quarkonium bound states. 

These qualitative considerations, however, illustrate already the difficulty 
that arises when 
one wants to use the concept of quarkonium suppression as a quantitative 
tool for studies of the QCD transition. On the one hand, it is the 
long distance property of QCD which undoubtedly will prohibit the 
formation of heavy quark bound states at high temperature. On the other 
hand, as heavy quark bound states are small in size also on the typical 
scale of QCD, $1/\Lambda_{QCD}\sim 1$fm, screening has to be 
strong enough to modify also the short distance part of the QCD potential. 
Already in the simple context of equilibrium thermodynamics different
length scales thus play a role which have to be understood quantitatively
before quarkonium suppression can be established as a unique signature for 
the existence of a phase transition or at least for the characterization 
of the properties of the hot and dense high temperature phase; a difficulty 
which is common to many signatures for the QCD transition that are currently 
discussed. 

A second ingredient to the concept of quarkonium suppression in heavy
ion collisions which also has been utilized in \cite{Matsui} is that the 
creation of a heavy quark pair in these collisions is a rare event; it  
is assumed that a $q\bar{q}$-pair that is created in a dense medium and
thus cannot form a bound state will separate unhampered and thereafter
both quarks will no more find 
another heavy quark as  partner to form a quarkonium bound state during 
the stage of hadronization.
While this was a good approximation for the kinematic conditions at 
the SPS it might not be appropriate anymore at RHIC or LHC. Consequently it 
recently has been suggested that recombination processes \cite{Thews,Rapp}
during the cooling of a hot plasma and subsequent pairing of heavy 
quarks initially originating from different creation processes (statistical
hadronization \cite{pbm,Rapp2}) may lead to a different suppression pattern or 
even to enhanced yields of heavy quark bound states. 

\begin{figure*}
\begin{center}
\epsfig{file=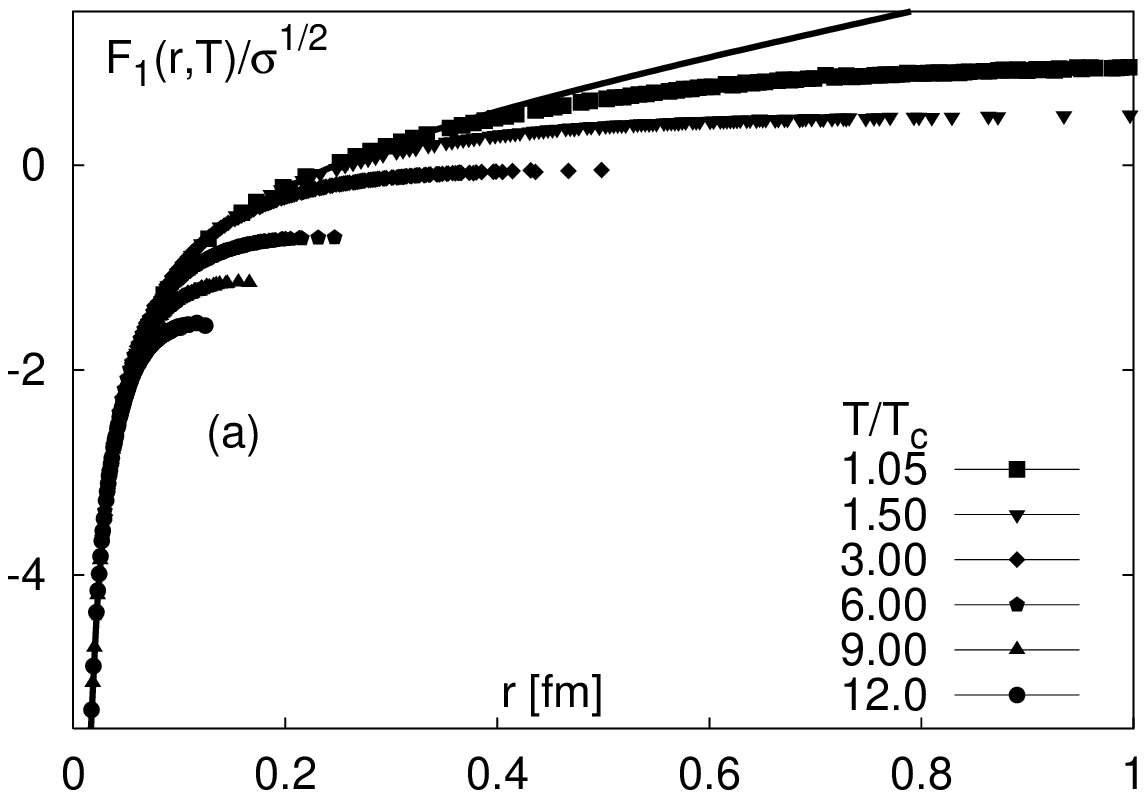,width=8.0cm}
\epsfig{file=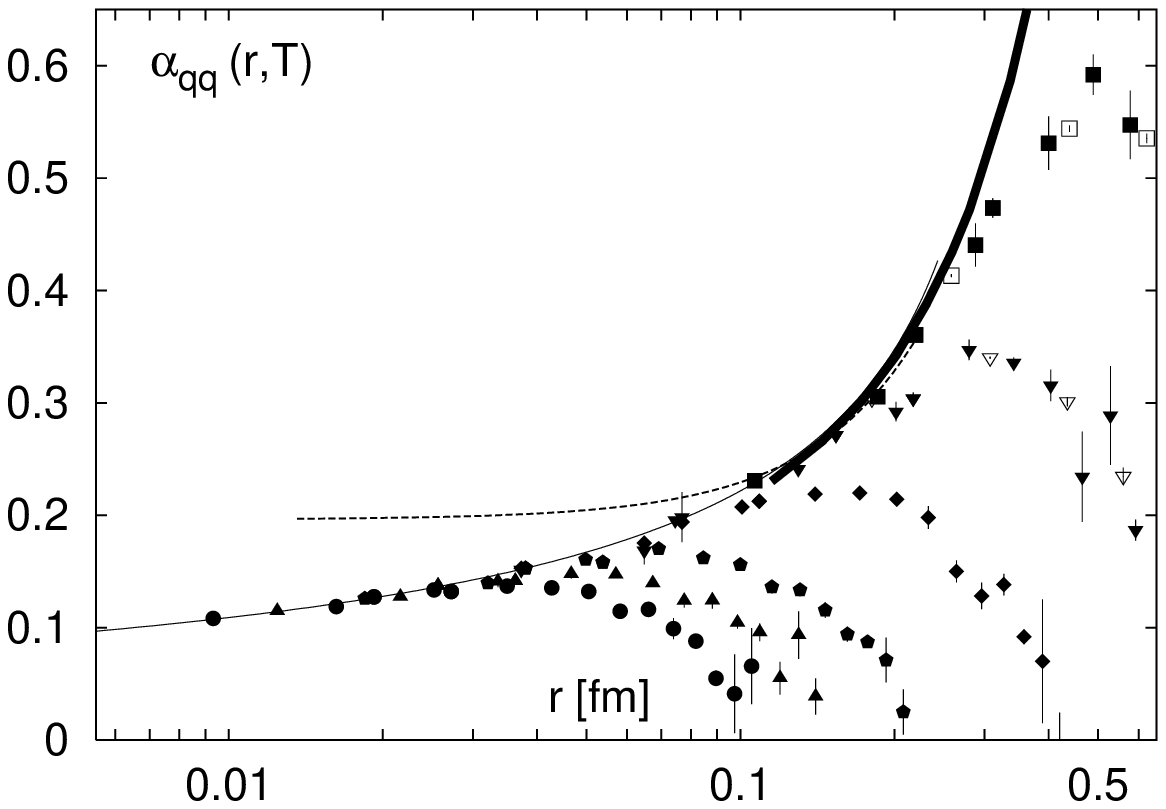,width=8.0cm}
\end{center}
\caption{Singlet heavy quark free energies (left) and the running coupling
in the so-called $qq$-scheme (right) for temperatures above the transition
temperature $T_c$. Results shown are obtained from
calculations in pure SU(3) gauge theory \cite{running}.}
\label{fig:free}     
\end{figure*}

A quantitative understanding of the heavy quark potential at finite 
temperature, the relevant lengths scales that characterize its short and long
distance properties and the in-medium properties of bound states combined 
with a thorough analysis of the time evolution of dense matter created in 
heavy ion collisions thus is needed to convert experimentally measured heavy 
quarkonium yields into a theoretically well understood tool that can be used 
for the detailed tomography\footnote{to use a modern buzzword in heavy ion 
phenomenology} of hot and dense matter. 
In fact, right from the beginning this was understood and has led to the 
discussion of sequential suppression patterns for bound states with
different quantum numbers \cite{Mehr}, its momentum and size 
dependence \cite{Petronzio} as well as the role 
of competing suppression mechanisms in ordinary hadronic matter \cite{Harris}.
All this has been discussed in detail in a recent report on the future heavy 
ion program at the LHC \cite{YB}.

We will concentrate here on a discussion of recent lattice calculations 
which aim at a better understanding of quarkonium properties at finite 
temperature. Through the detailed study of
short, medium and long-distance properties of heavy quark free energies
on the one hand and through calculations of spectral functions from thermal 
quarkonium correlation functions on the other hand much 
better inside into the fate of quarkonium states at high temperature
has been obtained.
In particular, a consistent picture now seems to emerge from potential
model calculations, based on calculations of heavy quark free energies,
and the direct spectral analysis of thermal hadron correlation functions.

We will start in the next section by discussing information on 
lengths scales relevant for the analysis of quarkonium bound states.
Section 3 contains a discussion of the calculation of effective 
potentials from heavy quark free energies and their role in
potential models. In Section 4 we present results on quarkonium
spectral functions obtained through a maximum entropy analysis of
thermal hadron correlation functions. A brief discussion of the
consequences of these results for the analysis of quarkonium 
yields in heavy ion collisions is given in Section 5. Finally
we give our Conclusions in Section 6.

\section{Short and long distance scales at finite temperature}
\label{sec:2}

A quite common qualitative argument for the dissolution of heavy quark bound 
states
is that these can no longer exist in a hot and dense medium once the screening
radius, {\it i.e.} the inverse of the (Debye) screening mass $r_D\equiv 1/m_D$,
becomes smaller than the typical size of a quarkonium bound state. To go
beyond such qualitative arguments it is necessary to analyze more 
quantitatively what small and large distances mean in quarkonium 
physics at a given temperature. 
Some insight into this question can be gained by analyzing
the short distance properties of excess free energies of a static 
quark anti-quark pair in a thermal medium \cite{McLerran}
\begin{eqnarray}
{\rm e}^{-\; F_{\rm av}(r,T)} &=& \frac{Z_{\bar{q}q} (T,V)}{Z (T,V)}
\nonumber  \\
&\equiv&
\frac{1}{V} \sum_{\vec{x},\vec{y}; |\vec{x}-\vec{y}|=r}
\frac{1}{9}
\langle {\rm Tr} L(\vec{x}){\rm Tr}L^{\dagger}(\vec{y})\rangle \quad ,
\label{freedef}
\end{eqnarray}
or the related singlet free energy, which is obtained from static
quark anti-quark sources with fixed color orientation corresponding to 
a color singlet state,
\begin{equation}
{\rm e}^{-\; F_{1}(r,T)} =
\frac{1}{V} \sum_{\vec{x},\vec{y}; |\vec{x}-\vec{y}|=r}
\frac{1}{3}\langle {\rm Tr}L(\vec{x})L^{\dagger}(\vec{y})\rangle_{gf} \quad .
\label{freedef_1}
\end{equation} 
Here $L(\vec{x}$ denotes the Polyakov loop constructed in lattice
calculations from a product of SU(3) matrices that represent the gluon
fields \cite{review}. In Eq.~\ref{freedef_1} the subscript, $gf$, indicates
that the expectation value has to be evaluated in a fixed gauge as the 
operator used to define $F_1$ is not gauge invariant. The corresponding
gauge invariant operator, for instance in Coulomb gauge, would be a 
complicated non local object \cite{Philipsen}.

The singlet free energy calculated in Coulomb gauge
is shown in Fig.~\ref{fig:free}(left) for a 
selected set of temperatures above the transition temperature. The results 
shown
are obtained from a calculation in a pure SU(3) gauge theory \cite{running}.
Recent results obtained in 2-flavor \cite{Zantow_nf2} and 3-flavor \cite{Petrov}
QCD are, however, also on a quantitative level quite similar. As can be 
seen, $F_1(r,T)$ coincides with the zero temperature heavy quark potential,
\begin{equation}
V(r) = - \frac{4}{3} \frac{\alpha}{r} + \sigma r \quad ,
\label{Cornell}
\end{equation} 
for a certain range of quark anti-quark separations and then turns over quite 
rapidly into a constant that decreases with temperature. In fact, 
these features become much more 
transparent from a calculation of the running coupling as a function 
of distance and temperature \cite{running,Necco}. It can conveniently 
be defined from a derivative of the singlet free energy ($T=0$: force)
which eliminates otherwise uncontrolled constants, 
\begin{equation}
\alpha_{\rm qq} (r,T) = \frac{3r^2}{4} \frac{{\rm d} F_1(r,T)}{{\rm d}r}
\quad .
\label{alpha}
\end{equation} 
This is shown in Fig.~\ref{fig:free}(right). At zero temperature the Cornell
type potential, Eq.~\ref{Cornell}, yields $\alpha_{\rm qq} (r,0) = \alpha
+ 3 \sigma r^2/4$, with $\alpha=\pi/16$ arising in string models as 
correction to the linear rising string tension term. This is
represented in Fig.~\ref{fig:free} by the dashed 
line at short distances and the fat solid line at large distances. 
Within this parametrization of the potential the entire $r$-dependence of 
the coupling arises
from the long distance confinement part of the potential. Replacing the
constant coupling in the Cornell potential by the perturbatively running
QCD coupling leads to a characteristic $r$-dependence also at short 
distances which is indicated by the thin solid line Fig.~\ref{fig:free}.  
It generally is argued that at $T=0$ the perturbative calculation of
the coupling in the so-called $qq$-scheme (Eq.~\ref{alpha}) 
is under control for $r\lsim 0.1$fm
\cite{Necco,pertrun}.

At short distances the running coupling at finite temperature coincides
with the $T=0$ coupling and, of course, runs with the dominant scale given
by the quark anti-quark separation,
\begin{equation}
\alpha_{\rm qq} (r,T) \sim \frac{6\pi}{(33-2 N_f)\ln(1/r\Lambda)}~,~{\rm for}~ 
r\Lambda \ll 1
~.
\label{alpha_short}
\end{equation}
At large distances, however, the coupling defined through Eq.~\ref{alpha} 
drops exponentially,
\begin{equation}
\alpha_{\rm qq} (r,T) \sim \alpha(T) \exp\{-m_D(T)\ r\}~,~{\rm for}~
r\gg 1/T~,
\label{alpha_large}
\end{equation}
which is in agreement with high temperature perturbation theory in a regime 
where temperature is the dominant scale. 

\begin{figure}
\begin{center}
\epsfig{file=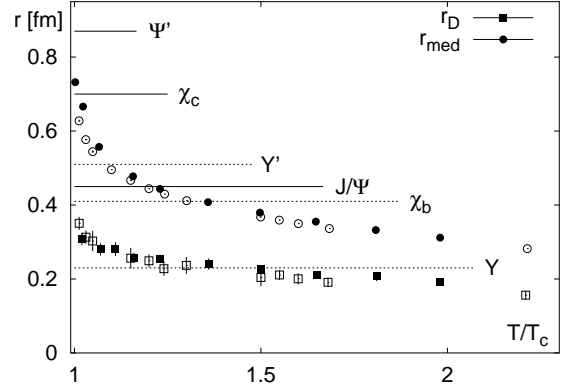,width=8.0cm}
\end{center}
\caption{The scale $r_{med}$ which gives an estimate for the distance beyond
which the force between a static quark anti-quark pair is strongly modified
by temperature effects and the Debye screening radius, $r_D\equiv 1/m_D$.
Open (closed) symbols correspond to SU(3) (2-flavor QCD) calculations.
The horizontal lines give the mean squared charge radii of some charmonium
and bottomonium states. }
\label{fig:scales}
\end{figure}

Similar results as those shown in Fig.~\ref{fig:free}(left) for the running
coupling of the pure gauge theory have been obtained also in 2 and 3 
flavor QCD \cite{Zantow_nf2,Petrov,Zantow_proc}. They clearly show that
(i) the coupling stays temperature independent in the short distance 
perturbative regime for quite a wide range of temperatures above $T_c$
and (ii) it does so also in the non-perturbative regime for temperatures 
up to about $1.5T_c$; {\it i.e.} the coupling  $\alpha_{\rm qq} (r,T)$
continues to show remnants of the confining force up to these temperatures. 
Moreover, (iii) the transition to the expected large distance behavior, 
Eq.~\ref{alpha_large}, proceeds quite rapidly
at a well defined distance, $r_{max}$, which can be attributed to the
point where $\alpha_{qq}(r,T)$ reaches a maximum at fixed $T$.
Due to the rapid crossover from short to large distance behavior 
above $T_c$ this length scale is, in fact, well be approximated by
the scale $r_{med}$ \cite{Zantow_nf2,Kaczmarek} defined as the point at which
$F_{\infty}\equiv \lim_{r\rightarrow \infty} F_1(r,T)$ equals the
zero temperature potential $V(r)$. 
This scale, extracted in a pure gauge theory \cite{Kaczmarek} and 
2-flavor QCD \cite{Zantow_nf2}, is shown in 
Fig~\ref{fig:scales} together with the scale set by the inverse Debye mass, 
$r_D\equiv 1/m_D$, which is extracted from the long-distance exponential 
screening of the singlet free energy.
Also included in this figure as horizontal lines are mean squared charge 
radii of some charmonium (solid) and bottomonium (dashed) states 
which characterize the average separation $r$ entering the effective potential
in the Schr\"odinger equation. It is reasonable to expect that
the temperatures at which these radii equal $ r_{med}$ provides a first 
estimate for the onset of thermal effects in quarkonium states. Of course, as
the wave functions of the various quarkonium states do also reach out to 
larger distances \cite{Jacobs} this can only be taken as a first indication
for the relevant temperatures. The analysis of bound states using, 
for instance, the Schr\"odinger equation will do better in this respect.  

\begin{figure*}
\begin{center}
\epsfig{file=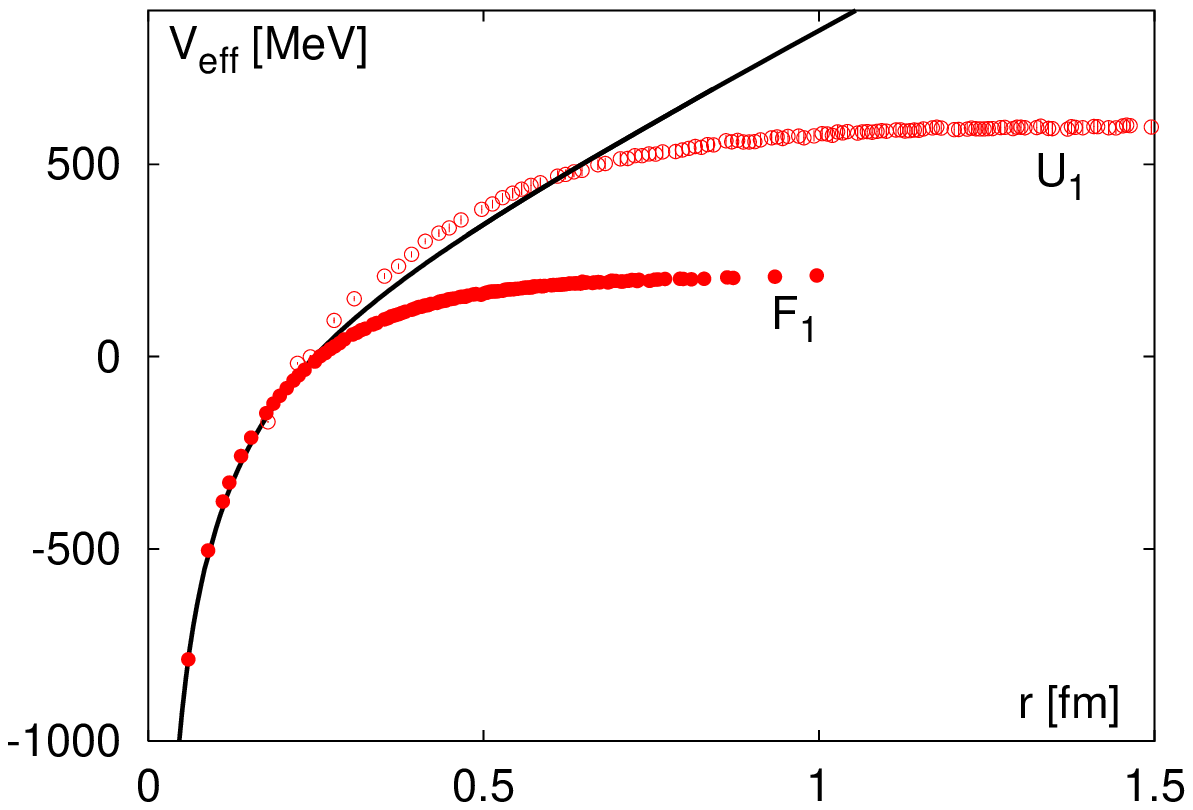,width=8.0cm}
\epsfig{file=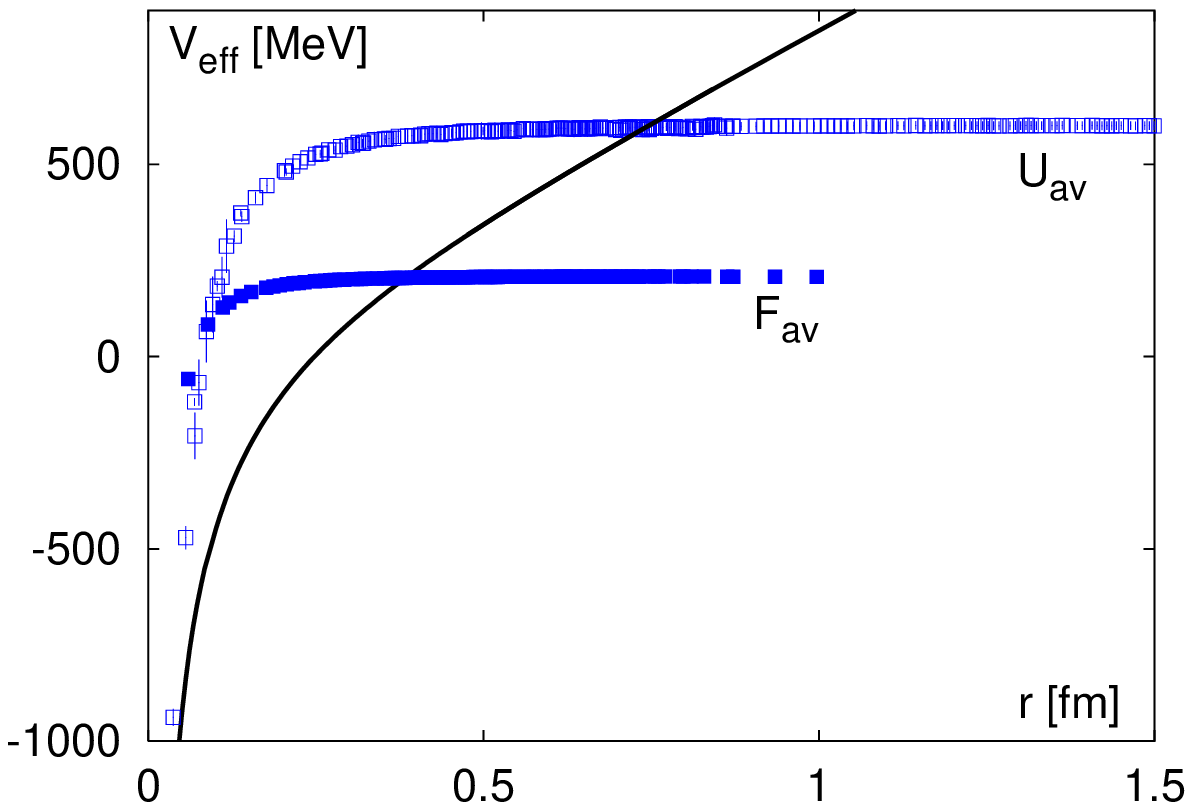,width=8.0cm}
\end{center}
\caption{Heavy quark free energies and energies in the color averaged
and singlet channels at $T\simeq 1.4 T_c$. Results shown are obtained from
calculations in pure SU(3) gauge theory \cite{Kaczmarek}.}
\label{fig:demo}     
\end{figure*}

\section{Non-Relativistic Bound State Models and the Heavy Quark Free Energies}
\label{sec:1}
The influence of a thermal medium on heavy quark bound states is commonly
discussed in terms of a non-relativistic, effective Hamiltonian for the 
quark anti-quark system \cite{Mehr,Digal,Wong}. 
Here the influence of the thermal medium is 
expressed in terms of a
temperature dependent potential for the two-quark system which then
can be used to study the bound state problem, for instance, by solving 
the corresponding non-relativistic Schr\"odinger equation,
\begin{equation}
\left[ 2m_f + {1\over m_f}\nabla^2 + {V_{eff}(r,T)} \right] \Phi_i^f =
{E_i^f(T)}\;
\Phi_i^f\quad,
\label{eq:schroedinger}
\end{equation}
where $m_f$ denotes the mass of the heavy quark flavor, e.g. $f\equiv$ charm or
bottom, $i$ labels different quantum number channels and $E_i^f$ is the 
corresponding binding energy, {\it i.e.} the bound state mass is then given 
by $M_i^f = 2 m_f+E_i^f$. Extensions of this approach taking into account
relativistic corrections have recently been considered in \cite{Shuryak}. 

The ''effective heavy quark potential'', $V_{eff}(r,T)$, 
models the influence of the medium on the $q\bar{q}$-system. This is then
used to solve the bound state problem for a two-body system only.
The main obstacle in such an approach
clearly is to obtain a suitable ansatz for $V_{eff}(r,T)$. Here lattice
calculations of the excess free energy of a pair of static quark anti-quark
sources in a thermal medium, defined in Eqs.~\ref{freedef} and \ref{freedef_1},
have frequently been used to define an effective potential. The often
used choice of the color averaged free energy as an effective potential,
$V_{eff}(r,T)\equiv F_{\rm av}(r,T)$, may be questionable
for two reasons. First of all  Eq.~\ref{eq:schroedinger} is meant
to solve the bound state problem for a color singlet $q\bar{q}$-state.
$F_{\rm av}(r,T)$, however, gives a weighted thermal average over 
$q\bar{q}$-states with arbitrary color orientation which can be 
understood in terms of, suitably defined, singlet and octet contributions.
One thus might want to use the free energy for a $q\bar{q}$-pair
fixed with relative color orientation corresponding to a color singlet 
state, $V_{eff}(r,T)\equiv F_1(r,T)$. 
Moreover, it might be more suitable to use as an effective potential
in Eq.~\ref{eq:schroedinger} the energy, $U(r,T)$, rather then the free 
energy, $F(r,T)$, of two heavy quarks in a thermal bath. 
The former can be obtained from calculations of the
free energy using the thermodynamic relation,
\begin{equation}
U(r,T) = -T^2\; \frac{\partial F (r,T)/T}{\partial T} \quad ,
\label{eq:energy}
\end{equation}
where $F$ and $U$ may correspond to the singlet or color
averaged free energy and energy, respectively.

Depending on the ansatz used for the effective potential the 
solution of the Schr\"odinger equation, Eq.~\ref{eq:schroedinger},
will lead to different dissociation temperatures. The qualitative
differences are obvious from Fig.~\ref{fig:demo} which shows for one
value of the temperature, $T=1.4\; T_c$, the
color averaged free energy and energy of static quark anti-quark sources (left)
as well as the corresponding quantities for the sources being fixed in
a relative singlet orientation (right). As the free energy includes an additional 
entropy contribution, $F=U-TS$, potentials defined in terms of the energy
are deeper than those defined in terms of free energies and thus will 
lead to higher dissociation temperatures. Likewise, the averaged (free)
energies also receive contributions from octet configurations, which
are repulsive at short distances. This leads to steeper (free) energies
than in the singlet channel. Dissociation temperatures of heavy quark
bound states determined from color averaged (free) energies are thus
generically smaller than the corresponding values obtained with effective
potentials defined in terms of singlet (free) energies. Some results 
for dissociation temperatures obtained by using as effective potential
singlet free energies \cite{Digal} and energies \cite{Wong}, respectively,
are summarized in Tab.~\ref{tab:td}. Both calculations are based on lattice
results obtained in the pure SU(3) gauge theory \cite{Kaczmarek}. 
Similar results have also been obtained in \cite{Shuryak}.
 
\begin{table*}
\begin{center}
\begin{tabular}{|c||c|c|c||c|c|c|c|c|}
\hline
 state & $J/\psi$ & $\chi_c$ & $\psi'$ & $\Upsilon$ & $\chi_b$ & $\Upsilon$' &
$\chi_b'$ & $\Upsilon$'' \\
\hline
$E_s^i$~[GeV] & 0.64 & 0.20 & 0.05 & 1.10 & 0.67 & 0.54 & 0.31 &
0.20 \\
\hline
{$T_d/T_c$~ [from~$F_1$]} & {1.1}  & {0.74} &
{0.1 - 0.2} & {2.31} & {1.13} &
{1.1}  & {0.83} & {0.74} \\
{$T_d/T_c$~ [from~$U_1$]} & {$\sim$ 2.0}  & {$\sim$ 1.1} &
{$\sim$ 1.1}  &
{$\sim$ 4.5} & {$\sim$ 2.0} & {$\sim$ 2.0}  &
{--} & {--} \\
\hline
\end{tabular}
\end{center}
\caption{\label{tab:td}Dissociation 
temperatures in the charmonium and bottomonium system
obtained by using pure gauge theory results for
singlet free energies \cite{Digal} (third row) and energies \cite{Wong}
(last row).
}
\end{table*}
For temperatures above the transition temperature
the free energies and energies that have been calculated now also
for QCD with dynamical quarks \cite{Zantow_nf2,Petrov} are quite similar
to the pure gauge theory results. The dissociation temperatures displayed
in Tab.~\ref{tab:td} thus give a good description of the current status
of potential model predictions. As may be obvious from Fig.~\ref{fig:demo}
the largest dissociation temperatures are obtained from $U_1$.

Despite the model dependence entering through the choice of $V_{eff}$
some generic features are common to all the potential model calculations.
Quite independently of the choice of $V_{eff}$ these calculations suggest 
that radial excitations in the charmonium
system ($\chi_c$, $\psi '$) cannot exist as bound states in the high
temperature phase. In fact, the disappearance of $\chi_c$ which will 
show up experimentally as a missing contribution to the $J/\psi$ yield
seems to be a good probe for the transition temperature. 
Calculations based on using the singlet energy as an effective potential,
furthermore, suggest that the ground states ($J/\psi$, $\eta_c$) may still 
survive in the high temperature phase to
temperatures as high as $2\ T_c$. This, however, is in contrast to findings
based on effective potentials constructed by using free energies rather 
than energies. This generically leads to smaller dissociation 
temperatures and, indeed, would suggest a dissociation of $J/\psi$ already 
at $T\simeq T_c$. 

So far we have argued in this section in favor of using color singlet
free energies (or energies) to discuss the fate of quarkonium bound states
in the context of potential models. Nonetheless, one may also put forward
arguments in
favor of an effective potential which incorporates some of the color
changing processes that take place in a thermal medium through the 
scattering of thermal gluons with the static sources.
In fact, such processes have been discussed as a mechanism for direct
quarkonium dissociation in a thermal medium \cite{Kharzeev}. The absorption
of  thermal gluons can change the color orientation of 
the quark anti-quark pair and put it into a color octet state. 
The analysis of color averaged (free) energies thus may have some 
justification also in the context of potential models.

\section{Heavy Quark Bound States from Spectral Analysis of Hadron
Correlation Functions}

It should have become clear from the discussion in the previous section
that an analysis of thermal properties of heavy quark bound states in
terms of potential models and probably even the determination of singlet
potentials from lattice calculations will require additional phenomenological
consideration. On the other hand, as pointed out in the Introduction,
an ab-initio approach to thermal hadron properties through lattice
calculations exists and is based on the calculation of thermal
hadron correlation functions,  
\begin{equation}
G_H(\tau,\vec{r}, T) = 
\langle J_H(\tau, \vec{r}) J_H^\dagger (0, \vec{0}) \rangle \quad ,
\label{def2pt}
\end{equation}
with $J_H=\bar{q}(\tau,\vec{r})\Gamma_H q(\tau,\vec{r})$ denoting a hadron 
current where the projection onto appropriate quantum numbers, $H$, 
is controlled by a suitable product of gamma matrices $\Gamma_H$. 
                                                                                
\begin{figure*}
\begin{center}
\hspace*{-2mm}
\epsfig{file=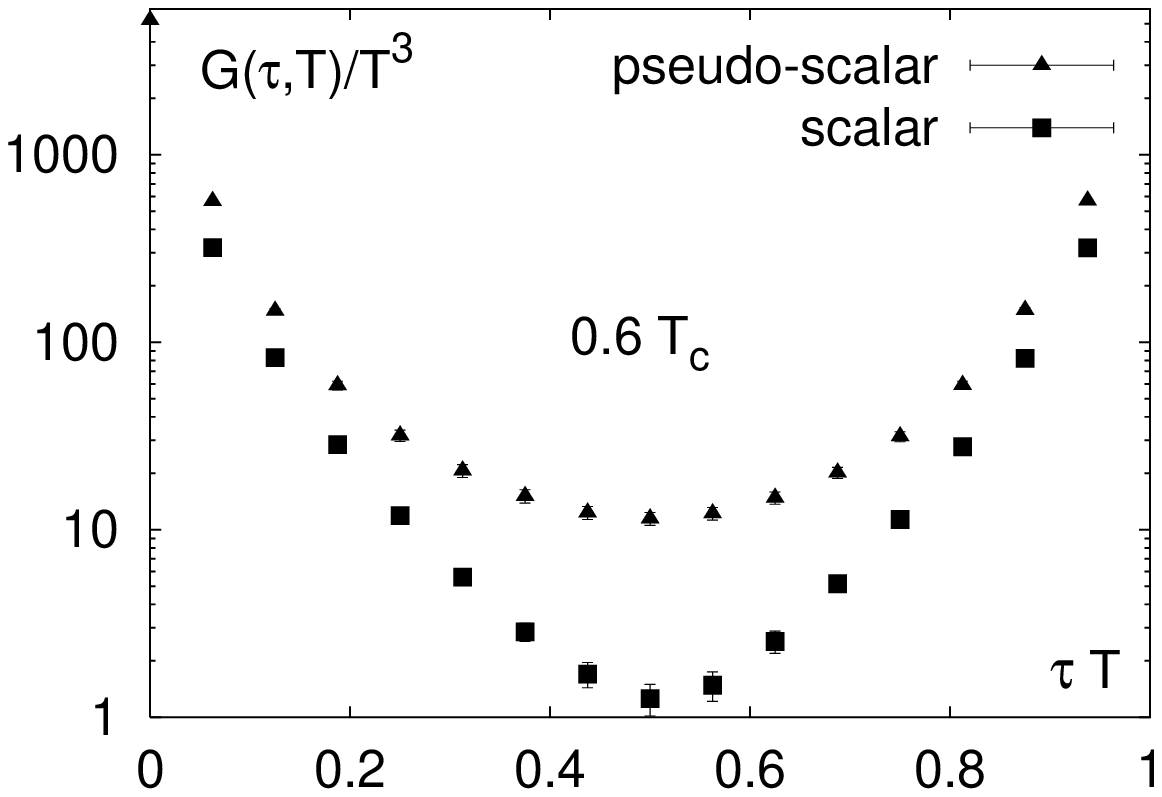,width=75mm}\hspace{0.2cm}
\epsfig{file=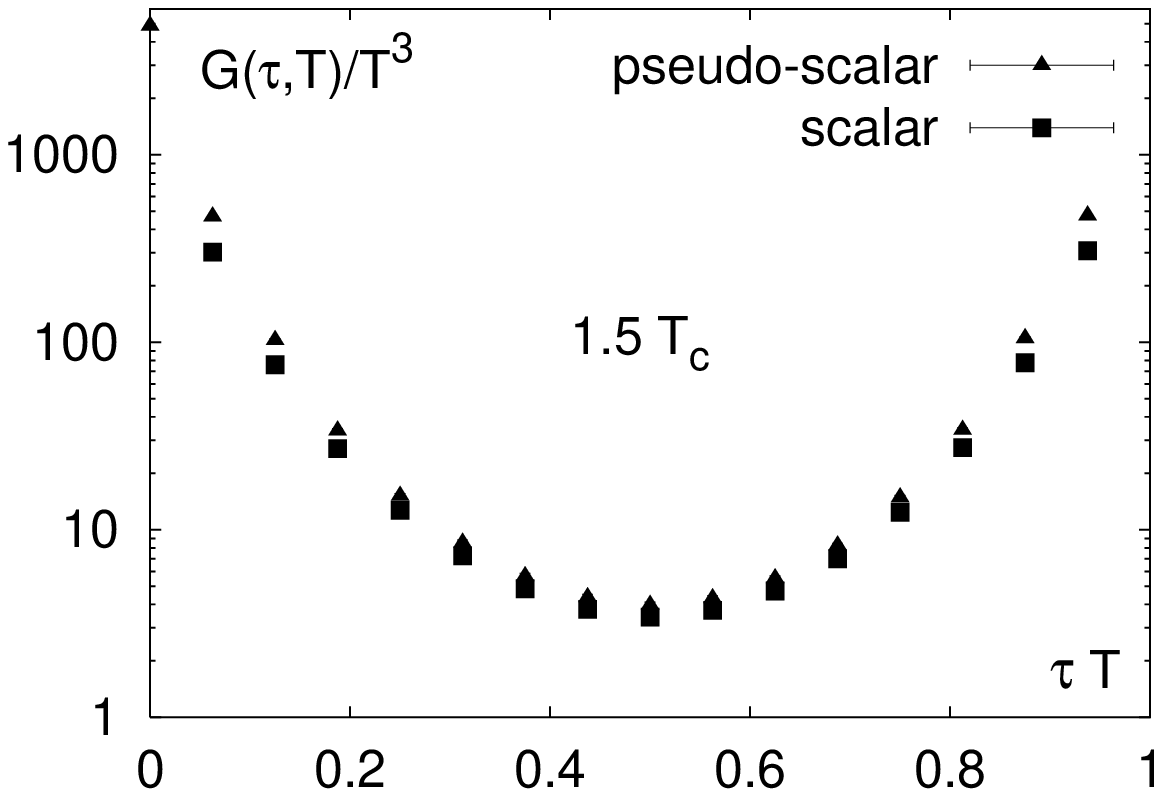,width=75mm}
\end{center}
\caption{\label{chiral} Scalar ($\delta$) and pseudo-scalar ($\pi$) correlation
functions at $T=0.6 T_c$ (left) and $T=1.5 T_c$ (right). Shown are results
from calculations with light quarks on a $64^3\times 16$ lattice in
quenched QCD.
}
\end{figure*}

The influence of a thermal heat bath on hadron properties immediately
becomes evident when one compares Euclidean time ($\tau$) correlation
functions of mesons in different quantum number channels. 
This is quite apparent for light quark mesons where thermal effects 
are related to the weakening of chiral symmetry breaking with increasing
temperature and its restoration at high temperature. To illustrate the
strong temperature effect on hadron correlation functions we thus 
briefly interrupt our discussion of heavy quark systems and have a look
at the light quark, chiral sector of QCD. 
At low temperature chiral symmetry
breaking and the breaking of the axial $U(1)$ symmetry lead to a splitting
of hadronic states, which would be degenerate otherwise and
would then also have identical correlation functions. In
Fig.~\ref{chiral} we show pseudo-scalar ($\pi$) and scalar ($\delta$) meson
correlation functions in a gluonic heat bath (quenched QCD) at temperatures
below (left) and above (right) the transition temperature.
The spectrum in the pseudo-scalar and
scalar channels differs at low temperature due to the explicit breaking
of the axial $U(1)$ symmetry which consequently leads to quite different
correlation functions in Fig.~\ref{chiral}(left). In fact, the difference
will increase with decreasing quark mass values as the Goldstone nature
of the pseudo-scalar than will lead to a flat correlation function.
Above $T_c$ 
the almost perfect degeneracy of both correlation functions  
suggests that the axial U(1) symmetry is effectively restored. Moreover,
the significant exponential drop of both correlation functions suggests
that both modes are ''quite heavy'' in the deconfined phase of QCD 
(Fig.~\ref{chiral}(right)).
                                                                                
The hadron correlation functions, $G_H(\tau,T)$, are directly related to
spectral functions, $\sigma_H(\omega, T)$, which contain all the information
on thermal modifications of the hadron spectrum in the quantum number  
channels, $H$,
\begin{eqnarray}
G_H (\tau,\vec{r},T) &&= \nonumber \\[1mm]
&& \hspace{-0.9cm}
\int_{0}^{\infty} {\rm d}\omega\; \frac{{\rm d}^3\vec{p}}{(2\pi)^3}
\sigma_H (\omega, \vec{p}, T) \;{\rm e}^{i\vec{p} \vec{r}}
{\cosh (\omega (\tau - 1/2T)) \over \sinh ( \omega /2T )}~ . \nonumber \\
~&&~
\label{correlator}
\end{eqnarray}
From the knowledge of e.g. the vector spectral function one 
obtains, for instance, direct information on thermal dilepton rates,

\begin{equation}
{{\rm d} W \over {\rm d}\omega {\rm d}^3p} =
{5 \alpha^2 \over 27 \pi^2} {\sigma_V(\omega,\vec{p},T)
\over \omega^2 ({\rm e}^{\omega/T} - 1)} \quad .
\label{rates}
\end{equation}
Of course, it should be clear that the rates calculated in this
way do not include any contributions arising from the 
feed-down of other quantum number channels into the
vector channel \cite{satz_feed}. 

Lattice studies of in-medium properties of hadrons have greatly advanced
in their predictive power through the exploitation of the Maximum
Entropy Method (MEM) \cite{Bryan,Hatsuda1}. This
allows the reconstruction of $\sigma_H (\omega,\vec{p}, T)$ at non-zero
temperature for light as well as heavy quark bound states.
We will concentrate in the following again on the heavy quark sector and, 
in particular,
will discuss in how far the potential model calculations presented in the
previous section are consistent with studies of charmonium spectral functions 
in the high temperature phase of QCD \cite{Asakawa1,Datta1,Umeda1}.
We will restrict this discussion to the zero momentum sector of the charmonium
system, $\sigma (\omega, T) \equiv \sigma_H(\omega, \vec{0}, T)$, although 
also some preliminary results at non-zero momentum have been presented 
recently \cite{Datta} and also some exploratory work for bottomonium
has been performed \cite{petrov_hp}.

\begin{figure}
\begin{center}
\hspace*{-2mm}
\epsfig{file=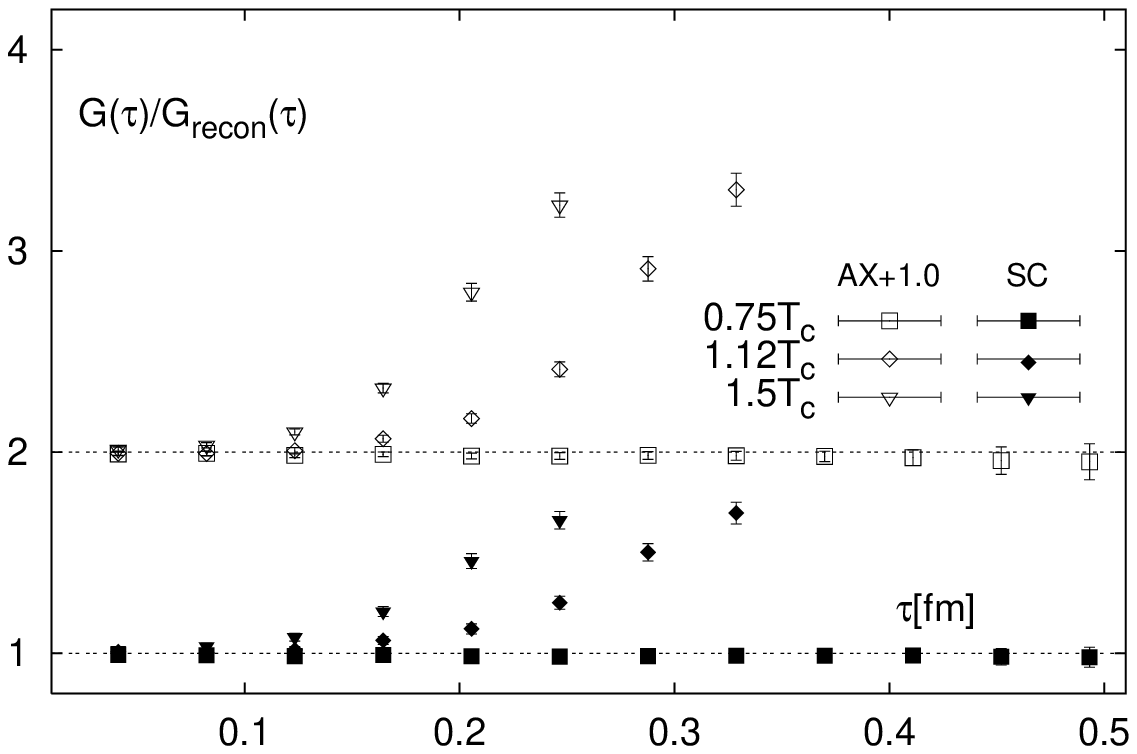,width=75mm}

\hspace{-0.2cm}\epsfig{file=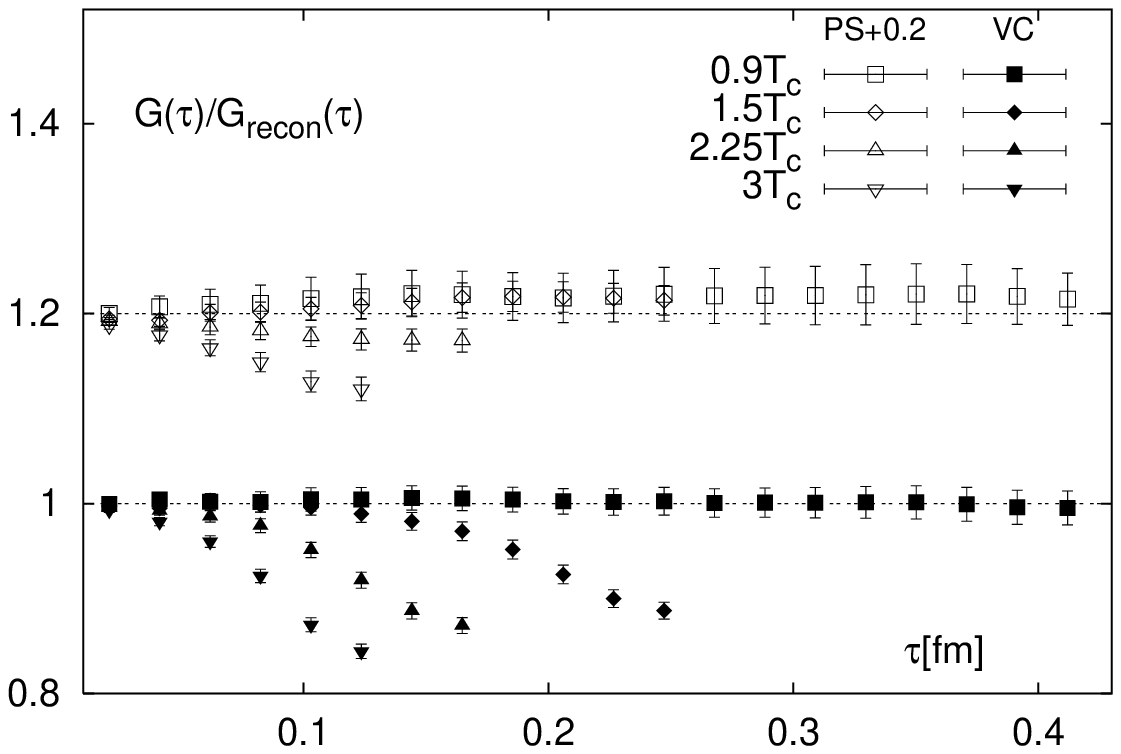,width=80mm}
\end{center}
\caption{\label{ratios}Thermal correlation functions above $T_c$ in axial
vector (AX: $\chi_{c,1}$) scalar (SC: $\chi_{c,0}$), 
pseudo-scalar (PS: $\eta_c$) and vector (VC: $J/\psi$) channels normalized
to reconstructed correlation functions which are based on spectral functions
calculated at $T^{*} = 0.75 T_c$ (top) and $T^{*} = 0.9 T_c$ (bottom) 
\cite{Datta1}. Note the different scales on the ordinates of both figures.}
\end{figure}

As indicated in Eq.~\ref{correlator}, $G_H(\tau,T) \sim \int {\rm d}^3\vec{r}
G_H(\tau,\vec{r},T)$ is related to the spectral functions 
$\sigma_H(\omega, T)$. However, an inversion of this integral equation,
which would be needed to extract $\sigma_H(\omega ,T)$ unambiguously, 
is generally not possible because
lattice calculations only yield information on $G_H(\tau,T)$ at a finite,
discrete set of Euclidean time steps, $\tau_k T = k/ N_\tau$ with $k=0, 1, ...
N_\tau -1$, with $N_\tau$ denoting the temporal extent of the lattice.
It is, however, possible to determine the {\it most probable} spectral
function which describes the calculated data set
$\{ G_H (\tau_k,T)\; | \; k=0, ..., N_\tau-1 \}$
and respects known constraints on $\sigma_H (\omega, T)$
(positivity, asymptotic behavior, ...).
This can be achieved using a Bayesian data analysis, e.g. the Maximum
Entropy Method (MEM) \cite{Bryan}. In the context of QCD calculations
on the lattice
this has been introduced in \cite{Hatsuda1}. 

The MEM analysis has its
own set of open problems. In particular, at finite temperature
some ambiguities arise through the sensitivity of details of the 
spectral functions to the input spectral function used as default model
to start the analysis \cite{Datta1,Asakawa_rev}. Moreover, also the high 
energy part of the spectral functions is strongly influenced by details of the
scheme used to discretized the lattice fermion action (Wilson doublers
\cite{CPPACS,Stickan}). This makes a ''point-by-point'' comparison of 
spectral functions determined by different groups difficult\footnote{In
any case, one should view the presently available set of calculations of 
spectral functions as exploratory leaving many possibilities for 
improvement.}.  

\begin{figure*}
\begin{center}
\setlength{\unitlength}{1.0cm}
\begin{picture}(10.0,8.0)
\boldmath
\put(4.3,5.11){\epsfig{bbllx=0,bblly=175,bburx=564,bbury=514,
file=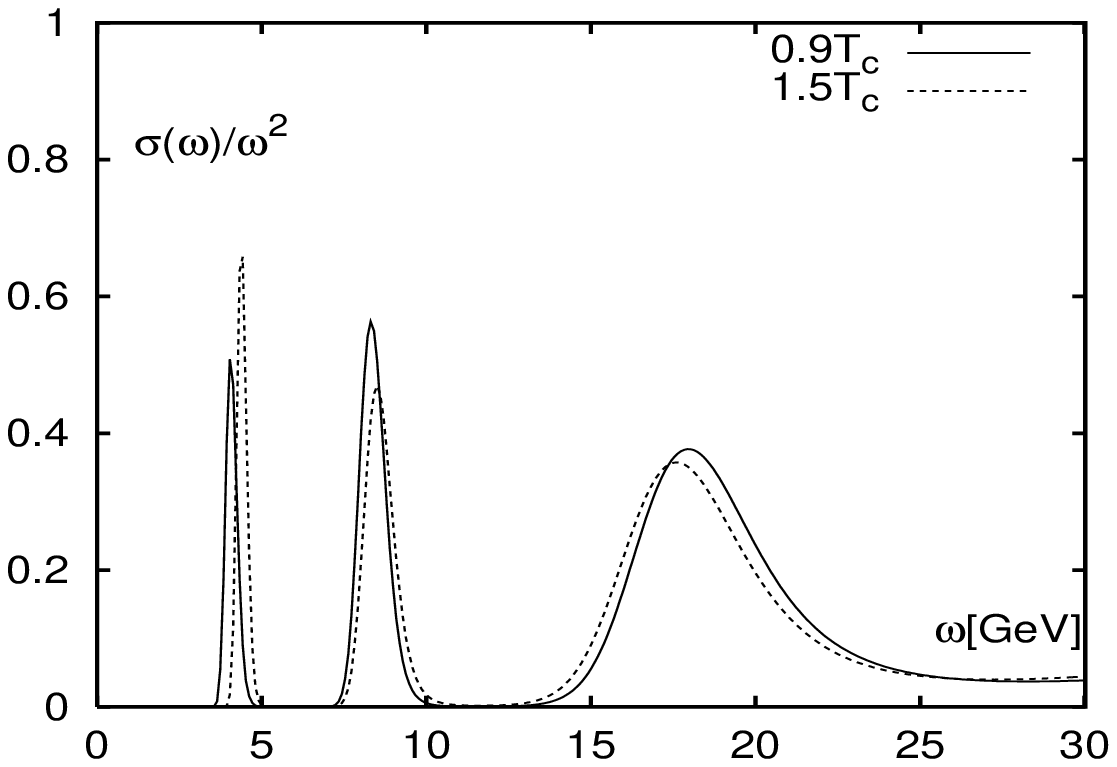,width=105mm}}
\put(4.2,2.77){\epsfig{file=whitebox2.eps,width=120mm,height=0.4cm}}
\put(4.3,1.41){\epsfig{bbllx=0,bblly=175,bburx=564,bbury=514,
file=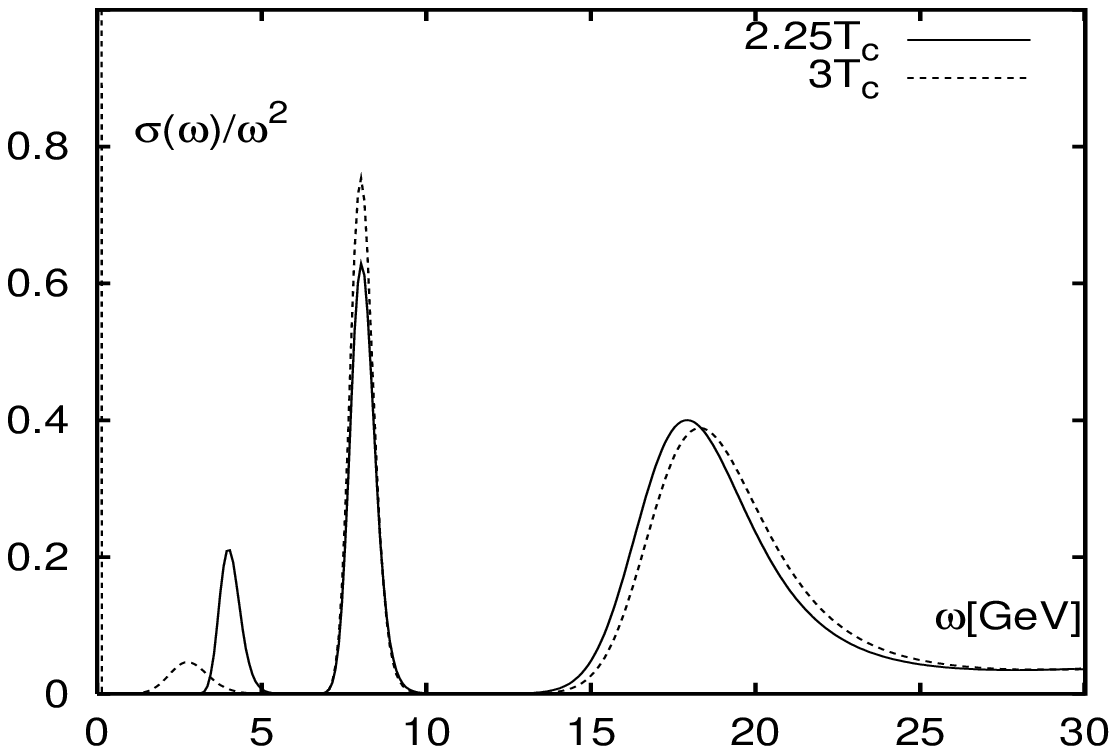,width=105mm}}
\put(-1.7,-1.0){\epsfig{bbllx=0,bblly=0,bburx=564,bbury=742,
file=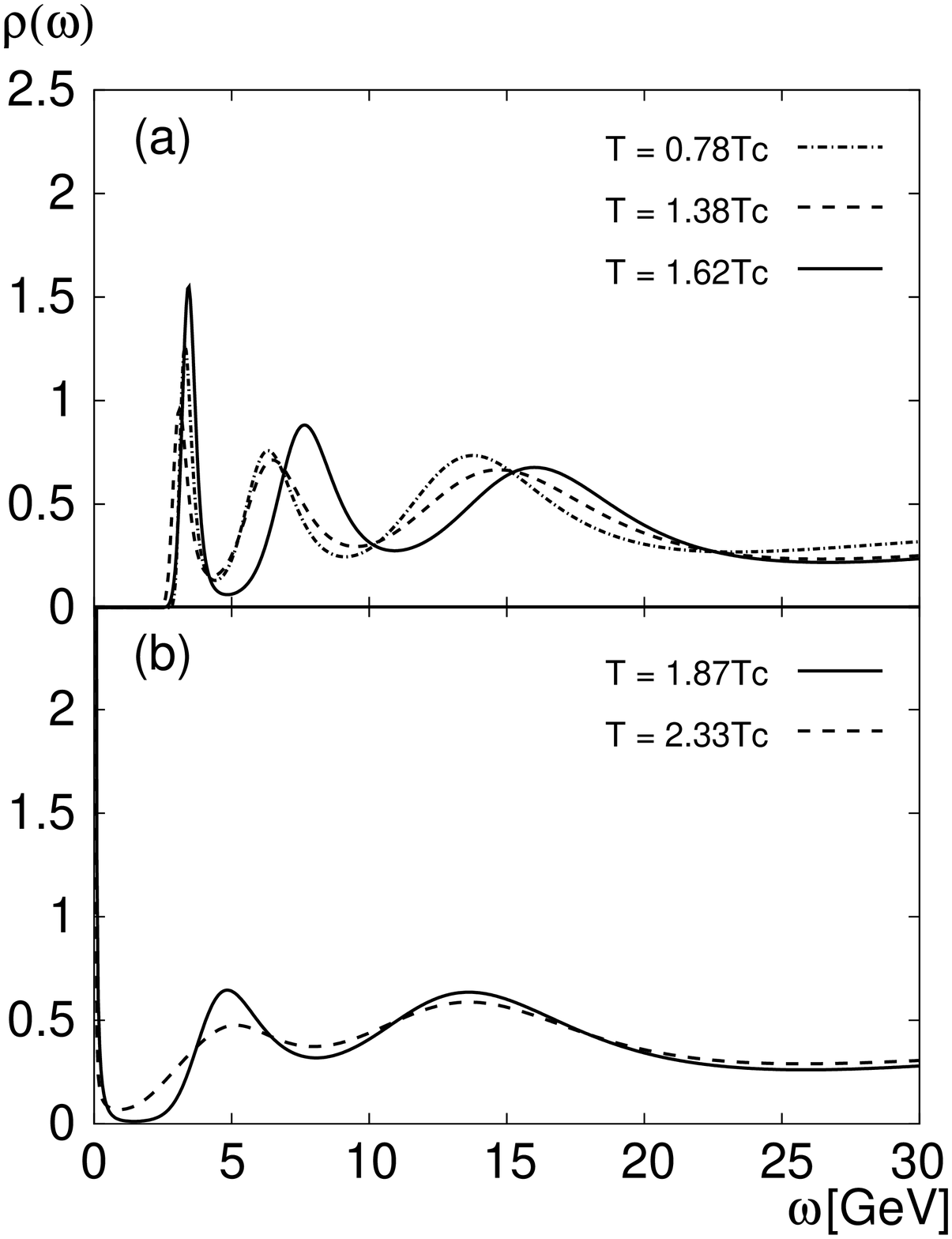,width=65mm}}
\put(-0.7,2.8){\epsfig{file=whitebox2.eps,width=0.6cm,height=0.4cm}}
\put(-0.7,6.3){\epsfig{file=whitebox2.eps,width=0.6cm,height=0.4cm}}
\put(-1.45,7.1){\epsfig{file=whitebox2.eps,width=0.7cm,height=0.4cm}}
\unboldmath
\end{picture}
\end{center}
\vspace*{0.5cm}
\caption{\label{spectral}Thermal vector spectral functions ($J/\psi$)
from MEM analyzes of meson correlation functions calculated in quenched
QCD on anisotropic \cite{Asakawa1} (left) and isotropic \cite{Datta1}
(right) lattices. The difference in scale on the ordinates of both figures
is due to different normalizations used for the hadronic currents.}
\end{figure*}

A first feeling for the influence of a thermal heat bath on the
structure of quarkonium correlation functions and the size of medium 
modifications of a thermal spectral function can be obtained
without relying entirely on the statistical MEM analysis
by comparing directly the numerically calculated correlation functions
at temperature $T$ with a correlation function, $G_{\rm recon} (\tau,T)$,
constructed from the spectral function calculated at a smaller (zero)
temperature, $T^{*} < T$, {\it i.e.}
\begin{equation}
G_{\rm recon} (\tau,T)
= \int_{0}^{\infty} {\rm d}\omega\;
\sigma_H (\omega, T^{*}) \;
{\cosh (\omega (\tau - 1/2T)) \over \sinh ( \omega /2T )} \quad .
\label{correlator2}
\end{equation}
In this way the trivial temperature dependence of the integration kernel
is taken care of and any remaining discrepancies between the reconstructed
correlation function and the actually calculated correlation function at
temperature $T$ can be attributed to changes in the spectral function.
Some results for charmonium correlation functions normalized in this way
are shown in Fig.~\ref{ratios}.
                                                                                
The ratio of correlation functions shown in Fig.~\ref{ratios} would equal 
unity, if the spectral functions would not depend on temperature. 
However, as can be seen
the correlation functions for $P$-state charmonium (top), show strong thermal
modifications already at temperatures slightly above $T_c$. On the
other hand, the $S$-state correlators (bottom) show only little modifications 
up to $T\simeq 1.5\; T_c$. In fact,
no temperature dependence is visible in the pseudo-scalar channel ($\eta_c$)
up to $T= 1.5\; T_c$ and also modifications in the vector channel ($J/\psi$) 
are only of the order of $10\%$ at this temperature.

The reconstruction of charmonium spectral functions has been performed
using technically different approaches, e.g. using point-like
\cite{Asakawa1,Datta1} or smeared \cite{Umeda2} hadron sources on
isotropic \cite{Datta1}
or anisotropic \cite{Asakawa1,Umeda1} lattice with standard Wilson
fermion \cite{Asakawa1} or improved Wilson fermion \cite{Datta1,Umeda1} 
actions. The different calculations agree to the extent that no significant
modification of $S$-state spectral functions is observed up to
$1.5\; T_c$, {\it i.e.} $J/\psi$ and $\eta_c$ survive as narrow bound 
states with
unchanged mass up to this temperature. This is shown in the upper part of
Fig.~\ref{spectral} for $J/\psi$. Current lattice calculations
\footnote{We stress again that current
lattice calculations of spectral functions are influenced by
lattice cut-off effects which show up most strongly at large energies,
$\omega$. In fact, only the first, low energy peak in the spectral functions
shown in Fig.~\ref{spectral} is physical and insensitive to changes of the
lattice cut-off. The other two peaks have been shown to be lattice
artifacts arising from ''Wilson doublers'' \cite{CPPACS}.}
differ, however, on the
structure of spectral functions for larger temperatures (lower part of
Fig.~\ref{spectral}). While it is concluded in \cite{Asakawa1} that
the $J/\psi$ resonance disappears quite abruptly at $T\simeq 1.9\; T_c$
the analysis of \cite{Datta1} suggests that the resonance disappears
gradually; a resonance peak with reduced strength is still visible at
$T= 2.25\; T_c$ and finally disappears completely at $T= 3\; T_c$.

For the radially excited states ($\chi_c$) the strong temperature depends seen 
already on the level of the correlation functions (Fig.~\ref{ratios}) 
also is apparent in the reconstructed spectral functions \cite{Datta1}. 
No evidence for bound states is found in these channels above the transition 
temperature. Although finite temperature
lattice calculations at present cannot resolve higher excited states in a 
given quantum number channel, e.g. the $\psi'$ in the vector channel, it 
seems likely that also these excitations will get dissolved at $T_c$ or even 
earlier. All this is consistent with the revised potential model
calculations in terms of a singlet free energy.
                                                                                
To get control over the detailed pattern of dissolution of
the heavy quark resonances clearly requires more refined studies. It also 
should be noted that so far all existing lattice studies have been performed
in the quenched approximation. Although virtual quark loops are not
expected to modify the qualitative picture obtained from these calculations
in the heavy quark sector of QCD one clearly has to understand their influence
on a quantitative level. As discussed in the previous section also the 
analysis of free energies in pure gauge theories and QCD with dynamical quarks  
suggests that dynamical quarks do not alter the picture gained from
quenched calculations significantly.
                                                                                
\section{Quarkonium suppression pattern in heavy ion collisions}

So far we have discussed possible modifications of quarkonium spectra
in equilibrated quark-gluon matter kept at fixed temperature. 
Which impact the basic results of the spectral analysis of charmonium,
{\it i.e.} (i) dissociation of excited states close to $T_c$, (ii)
persistence of $J/\psi$ at least up to $1.5T_c$, will have on  
observable charmonium yields in heavy ion collisions crucially depends on the 
fate of these states during the expansion of the hot and dense medium
created in these collisions. To this extent it is important to understand 
what will happen to a quark anti-quark pair after its initial production
in a hard collision during the subsequent expansion and cooling of the dense 
medium created around it. If the conditions in the medium are such
that the initially created pair cannot form a bound state and will
fly apart, it still may be the case that recombination processes play 
a significant role during the cooling of the hot medium \cite{Thews,Rapp}.
If the latter is the case a scenario in which the observed charmonium 
yields may be interpreted in terms of a statistical hadronization
model \cite{pbm,Rapp2} seems likely to be valid. 
Contrary to the characteristic sequential suppression pattern
\cite{Mehr,Digal} which would result from the standard screening scenario
for quarkonium suppression \cite{Matsui} this would not lead to any
threshold effects in the energy or temperature dependence of  
charmonium yields. In fact, it may well be that the latter scenario
is favored for the kinematic conditions at SPS while the 
statistical hadronization picture could become valid at RHIC energies
and even more so at the LHC. 
While at the SPS at best a single $c\bar{c}$-pair is created per collision
quite a few pairs per collision can be created at RHIC and well above
100 pairs could be created in central collisions at the LHC. Such an 
estimate, taken from \cite{Thews:2002,Thews_YB}, is displayed in 
Tab.~\ref{tab:yields}.  
\begin{table}
\begin{center}
\begin{tabular}{llll}
\hline\noalign{\smallskip}
~ & SPS & RHIC & LHC \\
\hline\noalign{\smallskip}
$\sqrt{s} ({\rm GeV})$ & 18 & 200 & 5500 \\[1mm]
$N_{c\bar{c}} $ & 0.2 & 10 & 200 \\[1mm]
$N_{ch}$ & 1350 & 3250 & 16500 \\
\noalign{\smallskip}\hline
\end{tabular}
\end{center}
\caption{\label{tab:yields}
Estimate of the number of $c\bar{c}$-pairs created
in central collisions at the SPS, RHIC and LHC \cite{Thews:2002}. The
estimate is based on production cross sections calculated in \cite{Vogt}.
Also given is the total number of charged particles created in these
collisions.}
\end{table}

As yields in recombination or hadronization 
models are proportional to $N_{c\bar{c}}^2$, charmonium yields 
grow much faster than the average charged particle multiplicities,
\begin{equation}
N_{J/\psi} \sim N_{c\bar{c}}^2 / N_{ch} \quad . 
\label{yields_had}
\end{equation}
At RHIC and LHC the charmonium yields thus could even increase rather than 
decrease irrespective of whether
the original $c\bar{c}$-pair was dissolved by screening or other 
mechanisms. 
 
\begin{figure}
\vspace{3.0cm}
\begin{center}
\hspace*{-17.0cm}\epsfig{
file=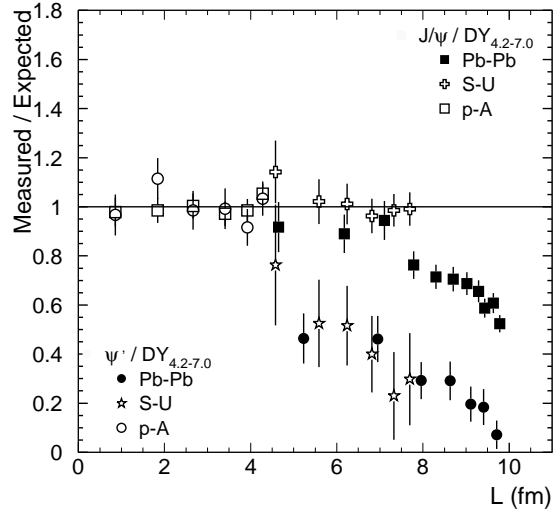,width=35mm}

\end{center}
\caption{\label{fig:suppression}
Experimental information on charmonium suppression at the SPS: Shown is 
the ratio ``measured value''/''expected value'' for the relative yields
$B_{\mu \mu}\sigma(J/\psi)/\sigma(DY)$ and $B_{\mu \mu}'\sigma(\psi')/\sigma
(DY)$ as a function of the nuclear absorption length $L$. The ''expected
value'' takes into account the ''normal'' nuclear absorption of charmonium
that traverses nuclear matter of thickness $L$.  
}
\end{figure}

If the formation of bound states is indeed controlled by 
the screening mechanism it may well be that the characteristic
sequential suppression pattern can only be established at the SPS
through a refined analysis of the charmonium system. 
The currently available data on $\psi'$ and $J/\psi$ yields from the SPS
\cite{NA50,NA50_last}, shown in Fig.~\ref{fig:suppression}, leave this 
possibility 
open. In fact, to substantiate the comparison of experimental results,
which are commonly expressed in terms of the nuclear absorption length $L$
\cite{NA50_last}, with expectations based on the screening mechanism it 
will also be important to reach a better understanding of the relation
between $L$, the energy density $\epsilon$ and the temperature $T$ of 
the dense matter created in a heavy ion collision. 
The relation between $L$ and $\epsilon$ is usually established through 
the Bjorken formula,
\begin{equation}
\epsilon = \frac{{\rm d} E_T /{\rm d}\eta}{\tau A_T} \quad ,
\label{Bjorken}
\end{equation}
where the geometrical overlap area $A_T$ is related to the impact 
parameter and can also be related to the absorption length $L$.  
Also the transverse energy per unit rapidity 
(${\rm d} E_T /{\rm d}\eta$), appearing in Eq.~\ref{Bjorken}, is 
experimentally accessible. Some uncertainty, however, is introduced 
through the choice of the formation 
time ($\tau$) used to characterize the initial state of the
medium at the time where $c\bar{c}$ bound states might form. This
generically is taken to be 1~fm. To further relate energy densities
to thermal properties of the medium it is necessary to know the
QCD equation of state which is taken from lattice calculations. 
In particular, the energy density at the transition point is presently
known only with large errors, $\epsilon_c = (0.3-1.3)$GeV/fm$^3$. 

Due to
the rapid change of energy density and pressure in the transition region
to the high temperature phase \cite{Peikert} small changes in the temperature
correspond to large changes in the energy density or $L$. For orientation
we list an estimate for the relation between these parameters in 
Tab.~\ref{tab:L} which is taken from fiures presented in \cite{NA50_last} 
and is based on an application of the Bjorken formula, Eq.~\ref{Bjorken}, 
with $\tau =1~$fm. The conversion
of this energy density to a temperature scale has then been performed
using the structure of the lattice equation of state by arbitrarily
fixing the energy density at $T_c$ to be 1~GeV/fm$^3$.
This shows that the experimentally observed suppression pattern may
actually correspond to a narrow temperature regime close to 
the transition temperature. 
 
\begin{table}
\begin{center}
\begin{tabular}{lll}
\hline\noalign{\smallskip}
$L$ [fm] & $\epsilon$ [GeV/fm$^3$] & $T/T_c$  \\
\hline\noalign{\smallskip}
4  & $\sim$1 & $\sim$1.0 \\[1mm]
8 &  $\sim$3 & $\sim$1.16  \\[1mm]
10 & $\sim$4 & $\sim$1.25 \\
\noalign{\smallskip}\hline
\end{tabular}
\end{center}
\caption{\label{tab:L}
Estimate of the relation between nuclear absorption length, energy
density and temperature in units of the transition temperature for
Pb-Pb collisions at the SPS assuming $\epsilon_c = 1$GeV/fm$^3$.} 
\end{table}

\section{Conclusions}

Quarkonium suppression has been observed in Pb-Pb collisions at the SPS.
The gross features are consistent with the original idea of $J/\psi$ 
suppression arising from screening of the heavy quark potential. 
It is conceivable that the suppression pattern observed at the SPS
is specific to the kinematic conditions in these experiments and 
that the enhanced creation of $c\bar{c}$ -pairs at RHIC and LHC may
also lead to an enhancement of $J/\psi$ yields at these colliders.

Lattice calculations for charmonium spectral functions as well as
refined potential model calculations suggest that the $J/\psi$ state
can survive up to temperatures well above $T_c$ whereas excited states
dissolve at or close to $T_c$. Given the current uncertainties in the
conversion of experimental estimates for the initial energy density
and theoretical relations between energy densities and temperatures
the observed suppression pattern thus is in line with the assumption
that it is caused by the dissolution of $\psi'$ and $\chi_c$ states.
At least on the theoretical side these uncertainties can and will be 
reduced through improvements in the determination of the transition 
temperature and the corresponding energy density.

\section*{Acknowledgments}
This work has partly been supported by BMBF under grant No.06BI102  
and funds of a Virtual Institute of the Helmholtz Association under
grant No. VH-VI-041.

\end{document}